\documentclass[superscriptaddress,twocolumn]{revtex4}
\usepackage[tbtags]{amsmath}
\usepackage{amsfonts}
\usepackage{mathrsfs}
\usepackage{amssymb}
\usepackage{graphicx}
\usepackage{epsfig,graphicx,times}
\usepackage{subfigure}
\usepackage{color}

\begin{document}

\title{High efficiency tomographic reconstruction of quantum states by quantum nondemolition measurements}
\author{J. S. Huang}
\affiliation{Quantum Optoelectronics Laboratory, School of Physics
and Technology, Southwest Jiaotong University, Chengdu 610031,
China}
\author{L. F. Wei\footnote{weilianfu@gmail.com}}
\affiliation{Quantum Optoelectronics Laboratory, School of Physics
and Technology, Southwest Jiaotong University, Chengdu 610031,
China} \affiliation{State Key Laboratory of Optoelectronic Materials
and Technologies, School of Physics and Engineering, Sun Yat-Sen
University Guangzhou 510275, China}
\author{C. H. Oh\footnote{phyohch@nus.edu.sg}}
\affiliation{Centre for Quantum Technologies and Department of
Physics, National University of Singapore, 3 Science Drive 2,
Singapore 117542, Singapore}

\begin{abstract}

We propose a high efficiency tomographic scheme to reconstruct an
unknown quantum state of the qubits by using a series of quantum
nondemolition (QND) measurements. The proposed QND measurements of
the qubits are implemented by probing the the stationary
transmissions of the dispersively-coupled resonator. It is shown
that only one kind of QND measurements is sufficient to determine
all the diagonal elements of the density matrix of the detected
quantum state. The remaining non-diagonal elements of the density
matrix can be determined by other spectral measurements by
beforehand transferring them to the diagonal locations using a
series of unitary operations. Compared with the pervious tomographic
reconstructions based on the usual destructively projective (DP)
measurements (wherein one kind of such measurements could only
determine one diagonal element of the density matrix), the present
approach exhibits significantly high efficiency for $N$-qubit
($N>1$). Specifically, our generic proposal is demonstrated by the
experimental circuit-quantum-electrodynamics (circuit-QED) systems
with a few Josephson charge qubits.

PACS number(s): 42.50.Pq, 03.65.Wj, 03.67.Lx, 85.25.Cp

\end{abstract}

\maketitle
\section{Introduction}

The reconstruction of an unknown quantum state from a suitable set
of measurements is called quantum tomography~\cite{Paris}, which is
a particularly important method in the study of quantum mechanics
and its various applications. Since the characterization of the
states is the central tasks in quantum-state engineerings and
controls, this technique is of great importance in the current
quantum information processing. Recently, many theoretical analysis
and experimental demonstrations have been devoted to implement the
desirable quantum-state tomographies for, e.g., the polarization
states of photons~\cite{White,Langford}, the electronic states of
trapped ions~\cite{H1}, and the solid-state qubits~\cite{Liu}, etc.
However, all these tomographic reconstructions are based on the
destructively-projective (DP) measurements, and thus are very
operational-complicated. This is because that each kind of DP
measurements, e.g., $\hat{P}_k=|k\rangle\langle k|$, is required to
be performed many times on many copies of the reconstructed state
for determining just one of the elements (e.g., $|c_k|^2$ in the
state $|\psi\rangle=\sum_kc_k|k\rangle$) in the density matrix
$\rho=|\psi\rangle\langle\psi|$.

Besides the usual DP measurements, quantum state could also be
detected by other strategies, typically such as the so-called
quantum nondemolition (QND) measurements. Basically, QND measurement
is a nondestructive detection, as the measurement-induced
back-action noises could be effectively suppressed by repeatedly
hiding them in certain observables which are not of interests. The
basic criteria for a QND measurement is that the repeated
measurements of an observable $\hat{o}$ of the same system should
yield the identical result. This means that the measured observable
must be commutative with the Hamiltonian $\hat{H}_{int}$ describing
the interaction between the measured system and the detector, i.e.,
$[\hat{H}_{int},\hat{o}]=0$. Historically, QND measurement is
proposed to explore the fundamental limitations of measurements, and
has been demonstrated in various branches of physics, such as in the
detection of gravitational waves~\cite{Braginsky}, quantum
optics~\cite{Pereira,Bencheikh,Grangier},
telecommunications~\cite{Levenson2}, and quantum
control~\cite{Wiseman}, etc. In recent years, the QND measurement
has also been successfully applied to probe the atomic qubits in the
cavity quantum electrodynamics (QED)~\cite{Nogues,Turchette}.
Furthermore, this technique was extensively used to the circuit QED
systems~\cite{Blais,Wallraff,Wallraff2,Wei,Gambetta,Bianchetti,Filipp}
for nondestructively reading out the superconducting qubits. This
QND measurement is implemented by measuring the transmission of the
driven microwave signals through a transmission line resonator,
which is dispersively coupled to the detected qubits. This is
because that the detected qubits can cause sufficiently large
state-dependent shifts of the resonator frequency. Thus by detecting
the signals of the shifted frequency of the resonator, the qubit
state will be read out. However, the QND measurements in the above
works~\cite{Blais,Wallraff,Wallraff2,Wei,Gambetta,Bianchetti,Filipp}
are only utilized to effectively distinguish the different logic
states of the detected qubit(s), which is (are) not prepared
initially at their superposed state.

Motivated by the above experiments, recently we proposed a new
scheme to nondestructively detect the superposition of these logic
states by the QND measurements~\cite{arXiv}. By taking account of
the full quantum correlations between the resonator and
dispersively-coupled qubit(s), our proposal shows that each detected
peak marks one of the logic states and the relative height of such a
peak is related to its corresponding superposed probability. This
means that one kind of the QND measurements can determine all the
diagonal elements of the density matrix of the measured quantum
state $\rho=|\psi\rangle\langle\psi|$. Similarly, the non-diagonal
elements of  $\rho$ could be determined by other kinds of QND
measurements by performing the suitable unitary operations to
transfer them into the measurable diagonal locations. Therefore, the
proposed tomographic reconstructing approach is high efficient for
$N$ ($N>1$) qubits, as the number of the kinds of the QND
measurements required is significantly decreased. For example, to
tomographically reconstruct a two-qubit state, the proposed $6$-kind
QND measurements are sufficient. This is significantly simpler than
the previous schemes (requiring $15$-kind measurements) based on
either the DPs~\cite{Liu} or the individual dispersive readouts of
the logic states~\cite{Filipp}.

The paper is organized as follows: Sec.~II gives our generic model
of the transmissions of a driven resonator. In Sec.~III, we provide
a detailed analysis of the QND measurement of a single-qubit state
by probing the transmissions of the driven resonator. Next, we show
how to use these QND measurements to tomographically reconstruct an
unknown single-qubit state in the experimental circuit-QED system.
The extensions to the two-qubit case are given in Sec.~IV, where the
advantage of our proposal (compared with the previous approach based
on the DP measurements) will be explicitly revealed. The possible
generalization to the $N$-qubit situation and summarizations of our
main results are finally given in Sec.~V.

\section{Transmission of a driven empty cavity}

For the detection of the states of the qubits, we investigate the
photon transmission of a driven resonator by studying the
steady-state properties of the resonator-qubits dynamics. For
generality, we consider a cavity-QED system consisting of $N$
qubits. The Hamiltonian reads
\begin{eqnarray}
H_{}=\hbar\omega_r\hat{a}^\dagger\hat{a}+\sum_{j=1}^N[\frac{\hbar\omega_{j}}{2}\sigma_{z_j}
+\hbar g_j(\sigma_{+_j}\hat{a}+\sigma_{-_j}\hat{a}^\dagger)],
\end{eqnarray}
where $a^{(\dagger)}$ and $\sigma_{\pm_ j}$ are ladder operators for
the photon field and the $j$th qubit respectively. Also, $\omega_r$
is the cavity frequency, $\omega_{j}$ the $j$th qubit transition
frequency, and $g_j$ the coupling strength between the $j$th qubit
and the resonator. Suppose that the cavity is coherently driven by
\begin{eqnarray}
H_d=\hbar\epsilon(\hat{a}^\dagger
e^{-i\omega_dt}+\hat{a}e^{i\omega_dt}),
\end{eqnarray}
where $\epsilon$ is the real amplitude and $\omega_d$ the frequency
of the applied external driving.

Under the Born-Markov approximation, the dynamics of the whole
system with the  dissipations and dephasings is described by the
following master equation~\cite{Walls}
\begin{eqnarray}
\dot{\varrho}_N&=&-\frac{i}{\hbar}[{H_N},\varrho_N]+\kappa \mathcal
{D}[a]\varrho_N+\sum_{j=1}^N\gamma_{1,j} \mathcal
{D}[\sigma_{-_j}]\varrho_N\nonumber\\
&&+\sum_{j=1}^N\frac{\gamma_{\phi,j}}{2}
\mathcal{D}[\sigma_{z_j}]\varrho_N,\\
&&H_N=H+H_d.\nonumber
\end{eqnarray}
Here, $\varrho_N$ is the density operator and the dissipation
superoperator is defined by $\mathcal
{D}[A]{\varrho_N}=A{\varrho_N}A^\dagger-A^\dagger
A{\varrho_N}/2-{\varrho_N}A^\dagger A/2$, which describes the
effects of the environment on the system. The parameters of the last
three terms in Eq.~(3) correspond to photon decay rate $\kappa$, the
$j$th qubit decay rate $\gamma_{1,j}$, and the $j$th qubit pure
dephasing rate $\gamma_{\phi,j}$, respectively.

In what follows, we begin with the master equation~(3) to calculate
the frequency-dependent transmission of the cavity, which is
proportional to the steady-state mean photon number
$\langle\hat{a}^\dagger\hat{a}\rangle$ in the cavity. Technically,
to satisfy the basic criteria for the desirable QND measurements of
the $N$-qubit system, we assume that the conditions
\begin{equation}
0<\frac{g_j}{\Delta_j},\,\frac{g_jg_{j'}}{\Delta_j\Delta_{jj'}},\,\frac{g_jg_{j'}}{\Delta_{j'}\Delta_{jj'}}\ll
1,\,\,j\neq j'=1,2,...,N,
\end{equation}
should be satisfied for assuring the effective dispersive coupling
$\sigma_{z_j}\hat{a}^\dagger\hat{a}$ between the $j$th qubit and the
cavity. These conditions assure also that the inter-bit interactions
are negligible. Above, $\Delta_j=\omega_j-\omega_r$ denotes the
detuning between the $j$th qubit and the cavity, and
$\Delta_{jj'}=\omega_j-\omega_j'$ the detuning between the $j$th and
$j'$th qubits.

For contrast, we first calculate the transmission spectrum of a
driven empty cavity. The Hamiltonian of the simplified system
reduces to ($\hbar=1$ throughout the paper)
\begin{eqnarray}
H_0=\omega_r\hat{a}^\dagger\hat{a}+\epsilon(\hat{a}^\dagger
e^{-i\omega_dt}+\hat{a} e^{i\omega_dt}).
\end{eqnarray}
After the time-dependent unitary transformation defined by the
operator $R=\exp(-i\omega_dt\hat{a}^\dagger\hat{a})$, we get the
effective Hamiltonian
 \begin{eqnarray}
\tilde{H}_0=R^\dagger H_0R-iR^\dagger\partial{R}/\partial{t}
=-\Delta_{dr}\hat{a}^\dagger\hat{a}+\epsilon(\hat{a}^\dagger+\hat{a}),
\end{eqnarray}
where $\Delta_{dr}=\omega_d-\omega_r$ is the detuning of the cavity
from the driving. Consequently, we get the master equation for such
a driven empty cavity
\begin{eqnarray}
\dot{\varrho}_0&=&-i[\tilde{H}_0,\varrho_0]
+\kappa(\hat{a}\varrho_0\hat{a}^\dagger-\hat{a}^\dagger\hat{a}\varrho_0/2-\varrho_0\hat{a}^\dagger\hat{a}/2),
\end{eqnarray}
where $\varrho_0$ is the density matrix of the empty cavity.

From the above master equation, one can easily obtain the equations
of motion for the expectation values of the relevant operators, such
as mean photon number inside the cavity
$\langle\hat{a}^\dagger\hat{a}\rangle=Tr(\hat{a}^\dagger\hat{a}\varrho_0)$:
\begin{subequations}
\label{eq:whole}
\begin{eqnarray}
\frac{d\langle\hat{a}^\dagger\hat{a}\rangle}{dt}&=
&-\kappa{\langle\hat{a}^\dagger\hat{a}\rangle}-2\epsilon\mathrm{Im}{\langle\hat{a}\rangle},\label{subeq:1}
\end{eqnarray}
with
\begin{eqnarray}
\frac{d\langle\hat{a}\rangle}{dt}&=&(i\Delta_{dr}-\frac{\kappa}{2}){\langle\hat{a}\rangle}-i\epsilon.\label{subeq:2}
\end{eqnarray}
\end{subequations}
The steady-state solution of Eq.~(8) gives
\begin{eqnarray}
\frac{\langle\hat{a}^\dagger\hat{a}\rangle_{ss}}{\epsilon^2}
=\frac{1}{(\omega_d-\omega_r)^2+(\frac{\kappa}{2})^2}.
\end{eqnarray}
Obviously, the transmission spectrum  of an empty cavity, which is
proportional to  $\langle\hat{a}^\dagger\hat{a}\rangle$, is
well-known Lorentzian: centered at $\omega_d=\omega_r$ with the
half-width $\kappa$. Certainly, when $\omega_d$ does not
sufficiently match the cavity frequency, no photon penetrates the
cavity and thus no transmission is recorded.

\section{Tomographic reconstruction of a single-qubit state by
QND measurements}

\subsection{Nondestructive detection of a single qubit by cavity
transmissions}

Now we investigate the case, in which a single qubit with transition
frequency $\omega_1$ is dispersively coupled to the cavity mode. In
the frame rotating at drive frequency $\omega_d$ characterized by
the transformation $R$, the Hamiltonian of the system reads
\begin{eqnarray}
\tilde{H}_{1}=\frac{\tilde{\omega}_1}{2}\sigma_{z_1}+(-\Delta_{dr}+\Gamma_1\sigma_{z_1})\hat{a}^\dagger\hat{a}
+\epsilon(\hat{a}^\dagger+\hat{a}),
\end{eqnarray}
with $\tilde{\omega}_1=\omega_1+\Gamma_1$ and
$\Gamma_1={g_1^2}/{\Delta_1}$.

Under the Born-Markov approximation, the master equation for the
single-qubit plus the driven resonator is
\begin{eqnarray}
\dot{\varrho}_1&=&-i[\tilde{H}_1,\varrho_1]+\kappa \mathcal
{D}[a]\varrho_1+\gamma_{1,1} \mathcal {D}[\sigma_{-_1}]\varrho_1
\nonumber\\&&+ \frac{\gamma_{\phi,1}}{2}
\mathcal{D}[\sigma_{z_1}]\varrho_1.
\end{eqnarray}
The desirable quantity $\langle\hat{a}^\dagger\hat{a}\rangle$ can be
determined by solving the following coupled equations of motion:
\begin{subequations}
\label{eq:whole}
\begin{eqnarray}
\frac{d\langle\hat{a}^\dagger\hat{a}\rangle}{dt}&
=&-\kappa{\langle\hat{a}^\dagger\hat{a}\rangle}-2\epsilon\mathrm{Im}{\langle\hat{a}\rangle},\label{subeq:1}
\end{eqnarray}
\begin{eqnarray}
\frac{d\langle\hat{a}\rangle}{dt}&
=&(i\Delta_{dr}-\frac{\kappa}{2}){\langle\hat{a}\rangle}-i\Gamma_1{\langle\hat{a}\sigma_{z_1}\rangle}-i\epsilon,\label{subeq:2}
\end{eqnarray}
\begin{eqnarray}
\frac{d\langle\hat{a}\sigma_{z_1}\rangle}{dt}
&=&(i\Delta_{dr}-\frac{\kappa}{2}-\gamma_{1,1}){\langle\hat{a}\sigma_{z_1}\rangle}-(i\Gamma_1+\gamma_{1,1}){\langle\hat{a}\rangle}
\nonumber\\&&-i\epsilon{\langle\sigma_{z_1}\rangle},\label{subeq:3}
\end{eqnarray}
and
\begin{eqnarray}
\frac{d\langle\sigma_{z_1}\rangle}{dt}&=&-\gamma_{1,1}({\langle\sigma_{z_1}\rangle}+1).\label{subeq:4}
\end{eqnarray}
\end{subequations}
It is obvious that the additional measurement-induced dephasing rate
$\gamma_{\phi,1}$ does not influence the solution of the equations.
As the decay $\gamma_{1,1}$ of the qubit is significantly less than
the decay rate $\kappa$ of the driven cavity, the average of
$\sigma_{z_1}$ could be safely assumed to be unchanged during the
measurement. In fact, the characterized time of the detection is
determined mainly by the decay of the cavity $\kappa$. Such a
quantity is about $2\pi\times1.69$ MHZ~\cite{Bianchetti}, which is
obviously larger than $\gamma_{1,1}=2\pi\times0.02$
MHZ~\cite{Wallraff2}. Experimentally, the time interval of
completing a single QND measurement is about
$T_e=40$ns~~\cite{Bianchetti}, this is significantly shorter than
the lifetime: $T_1\sim7.3$$\mu$s and the decoherence time:
$T_2\sim500$ns~\cite{Wallraff2}. Therefore, during such a readout
the decay of the qubit is negligible, i.e.,
$\langle\sigma_{z1}(T_e)\rangle=\exp(-\gamma_{1,1}T_e)(\langle\sigma_{z1}(0)\rangle+1)-1\approx
\langle\sigma_{z1}(0)\rangle$.

Under the steady-state condition, we obtain
\begin{eqnarray}
&&\frac{{\langle\hat{a}^\dagger\hat{a}\rangle}_{ss}}{\epsilon^2}\nonumber\\
&=&\frac{2}{\kappa}\times[(\frac{\kappa}{2}+\gamma_{1,1})(\frac{\kappa^2}{4}+\frac{\gamma_{1,1}\kappa}{2}+\Gamma_1^2-\Delta_{dr}^2)+\nonumber\\
&&(\Delta_{dr}+\Gamma_1\langle\sigma_{z_1}(0)\rangle)(\kappa\Delta_{dr}+\gamma_{1,1}\Delta_{dr}+\gamma_{1,1}\Gamma_1)]\nonumber\\
&&\times[(\frac{\kappa^2}{4}+\frac{\gamma_{1,1}\kappa}{2}+\Gamma_1^2-\Delta_{dr}^2)^2\nonumber\\
&&+(\kappa\Delta_{dr}+\gamma_{1,1}\Delta_{dr}+\gamma_{1,1}\Gamma_1)^2]^{-1},
\end{eqnarray}
which is strongly related to the the initial state of qubit, thus
the qubit state could be determined by the cavity transmission.

\begin{figure}[htbp]
\includegraphics[width=8cm,height=6cm]{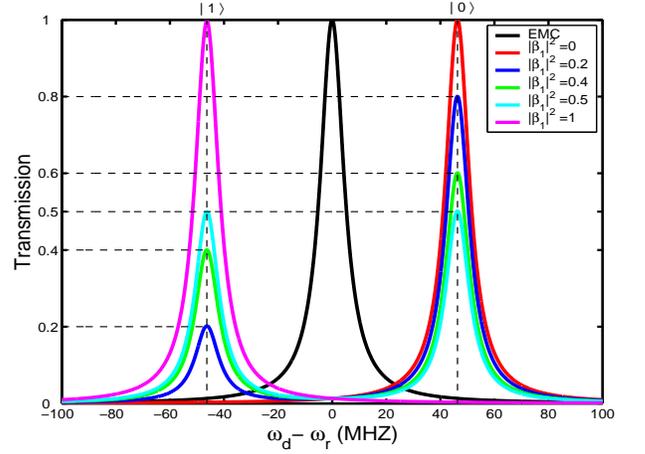}
\caption{(Color online) Cavity transmission for the single-qubit
states versus the probe detuning $\omega_d-\omega_r$.  Five cases of
the qubit states for $|\beta_1|^2=0, 0.2, 0.4, 0.5$, and $1$ are
shown. For comparison, the empty cavity (EMC) transmission is also
plotted in black line.
  The peak shifts  by
$-\Gamma_1$ or $\Gamma_1$ correspond to single logic state
$|0\rangle$ or $|1\rangle$. For the superposition states, the
double-peak relative heights
 (in contrast to the peak height of  the empty cavity
transmission)  present  clearly  the superposed probabilities of the
two logic states. Here, the parameters are selected as:
$(\Gamma_1,\kappa,\gamma_{1,1})=2\pi\times(-7.38,1.69,0.02)$MHz
~\cite{Bianchetti,Wallraff2}.}
\end{figure}

The measured cavity transmission (normalized to the peak height of
the empty cavity transmission) versus the probe frequency detuning
are plotted in Fig.~1. Generally, the qubit is assumed to be
prepared initially in the state
$|\psi\rangle=\beta_0|0\rangle+\beta_1|1\rangle$. Obviously, when
$\beta_0=0$, or $1$, it reduces to the single logic state
$|1\rangle$ or $|0\rangle$. Compared with the empty cavity
transmission (plotted as the dark line in Fig.~1), one observes that
qubit-resonator coupling leads to a right (left) shift of the single
peak in the transmission spectrum by a quantity $-\Gamma$
($\Gamma$), which is dependent of the logic states for
$\langle\sigma_{z_1}(0)\rangle=-1$
($\langle\sigma_{z_1}(0)\rangle=1$). Thus, the shifts of the peaks
can be used to mark the logic states of the qubit. However, when the
qubit is in the superposition of the two logic states, e.g.,
$|\beta_1|^2=0.2, 0.4$, and $0.5$, respectively, we see that the
situation is very different from the case for the single logic
states. In this case, the spectrum shows two peaks whose locations
coincide with that for the single logic states, but the relative
heights of these two peaks correspond clearly to the superposed
probabilities, i.e., $|\beta_0|^2$ and $|\beta_1|^2$, respectively.
This provides an effective approach to directly measure the
superposed probabilities of a superposed state.

\subsection{Tomographic reconstruction of a single-qubit state}

Above investigation indicates that, partial information of the qubit
state, i.e., the diagonal elements of the relevant density matrix,
can be directly obtained by only one kind of the QND measurements.
While, to extract the full information of an unknown qubit state,
one should tomographically reconstruct all the elements of its
density matrix. Basically, to completely define a $d$-dimensional
density matrix $\rho$, one needs to determine $d^2-1$ real
parameters. Therefore, to determine an unknown qubit state, the key
point is to identify these parameters by virtue of the tomographic
technique.

Now we demonstrate how to perform the tomographic construction of an
arbitrary single-qubit state
$|\psi\rangle_1=\beta_0|0\rangle+\beta_1|1\rangle$, whose density
matrix operator reads
\begin{eqnarray}
 \rho_1=\left(
\begin{array}{cccc}
\rho_{00} & \rho_{01}  \\
\rho_{10} & \rho_{11}  \\
\end{array}\right).
\end{eqnarray}
A more efficient and widely used technique is to parameterize the
density matrix $\rho_1$ on a Bloch sphere~\cite{Liu},
\begin{eqnarray}
 \rho_1=\frac{1}{2}(I
+\sum_{i=x,y,z}r_i\sigma_i) =\frac{1}{2}\left(
\begin{array}{cccc}
1+r_{z} & r_{x}-ir_{y}  \\
r_{x}+ir_{y} & 1-r_{z}  \\
\end{array}\right).
\end{eqnarray}
Here, $I$ denotes the identity matrix, $\sigma_i$ the Pauli
matrices, and $r_i$ real parameters. Therefore, in order to
determine  the single-qubit state, we must identify the three
components $(r_x, r_y, r_z)$ of the Bloch vector $\vec{r}$. As
discussed in the previous section, two diagonal elements $\rho_{00}$
and $\rho_{11}$ of the density matrix $\rho_{1}$ can be directly
determined by the two measured occupation probabilities
$|\beta_0|^2$, $|\beta_1|^2$ in the direct QND measurements. This
means that the parameter $r_z$ can be determined by the relation
$r_z=\rho_{00} -\rho_{11}$ $=|\beta_0|^2-|\beta_1|^2$. To obtain the
other two parameters $r_x$ and $r_y$, we need to determine the
non-diagonal elements. To this end, we perform the single-qubit
operations: $U_{x_1}=\exp{(i\pi\sigma_{x_1}/4)}$ and
$U_{y_1}=\exp{(i\pi\sigma_{y_1}/4)}$, to transfer them to the the
relevant diagonal locations, respectively. For example, after the
operation $U_{x_1}$, the density matrix $\rho_1$ is changed to
\begin{eqnarray}
 \rho_1'=U_{x_1}\rho_1 U_{x_1}^\dagger
=\frac{1}{2}\left(
\begin{array}{cccc}
1-r_{y} & r_{x}-ir_{z}  \\
r_{x}+ir_{z} & 1+r_{y}  \\
\end{array}\right).
\end{eqnarray}
Now, performing another kind of QND measurements the parameters
$|\beta'_0|^2$ and $|\beta'_1|^2$ can be measured. Consequently, the
coefficient $r_y$ can be determined via the relation
$r_y=|\beta'_1|^2-|\beta'_0|^2$. Similarly, by performing the
quantum operation $U_{y_1}$ on the original density matrix $\rho_1$,
another new density matrix
\begin{eqnarray}
 \rho_1''=U_{y_1}\rho_1 U_{y_1}^\dagger
=\frac{1}{2}\left(
\begin{array}{cccc}
1+r_{x} & -r_{z}-ir_{y}  \\
-r_{z}+ir_{y} & 1-r_{x}  \\
\end{array}\right),
\end{eqnarray}
can be obtained and the coefficient $r_x$ can be similarly
determined. Note that here the number of the unitary operations
required for implementing the quantum state tomography (based on the
QND measurements) is the same as the previous approach (based on the
usual DP measurements). Thus, for the single-qubit case the
complexity of the present approach is the same as that in the
previous one. Note that here, as the same as that in the previous
approach based on the DP measurements, three kinds of QND
measurements are still required for the present reconstructions. One
is directly applied, another is applied after the $U_{x_1}$
operation, and the final one is applied after the $U_{y_1}$
operation, thus the efficiency is not enhanced.

The remaining task is to implement the single-qubit operations
required above for transferring the non-diagonal elements to the
diagonal locations. We work with a circuit-QED system wherein a
superconducting charge qubit is coupled to the fundamental mode of a
transmission line resonator~\cite{Makhlin}. Let the qubit work at
its degeneracy point and neglect the fast oscillating terms under
the rotating-wave approximation (RWA). Following Ref.~\cite{Blais2},
under one displacement transformation, the effective Hamiltonian of
the resonator plus qubit system can be written as
\begin{eqnarray}
\tilde{H}=-\Delta_{dr}\hat{a}^\dagger\hat{a}+
\frac{\Delta_a}{2}\sigma_{z_1}
+g_1(\hat{a}^\dagger\sigma_{-_1}+\hat{a}\sigma_{+_1})
+\frac{\Omega}{2}\sigma_{x_1},\nonumber\\
\end{eqnarray}
with the detuning of the qubit transition frequency from the drive
$\Delta_a=\omega_1-\omega_d$ and the Rabi frequency
$\Omega=2\epsilon g_1/(-\Delta_{dr})$. Next, supposing that this
system works in the dispersive regime, i.e., $|g_1/\Delta_1|\ll 1$,
after the transformation
$U_1=\exp{[{-g_1}(\hat{a}^\dagger\sigma_{-_1}-\hat{a}\sigma_{+_1})/{\Delta_1}]}$,
then the above Hamiltonian becomes
\begin{eqnarray}
H_x =-\Delta_{dr}\hat{a}^\dagger\hat{a}
+\frac{\tilde{\Delta}_a}{2}\sigma_{z_1}+\frac{\Omega}{2}\sigma_{x_1},\,
\tilde{\Delta}_a=\Delta_a+\Gamma_1.\nonumber\\
\end{eqnarray}
First, if the condition $\tilde{\Delta}_a=0$ is satisfied, then the
Hamiltonian (19) produces a rotation of the qubit about the $x$
axis, i.e., $U_{x_1}$ could be generated by choosing the evolution
time $t_x=\pi/(2\Omega)$. Second, if the driving is sufficiently
detuned from the qubit and its amplitude is also sufficiently large
enough, then another approximate Hamiltonian
\begin{eqnarray}
&&H_z =-\Delta_{dr}\hat{a}^\dagger\hat{a}+
\frac{1}{2}(\tilde{\Delta}_a+\frac{1}{2}\frac{\Omega^2}{\Delta_a})\sigma_{z_1},
\end{eqnarray}
can be obtained by further performing a transformations
$U_2=\exp{({\beta}^*\sigma_{+_1}-{\beta}\sigma_{-_1})}$, with the
coefficient $\beta=\Omega/(2\Delta_a)$, on the Hamiltonian (19).
Obviously, the desirable operation $U_{z_1}$ can be implemented by
the evolution under the Hamiltonian (20) with the duration
$t_z=\pi\Delta_a/(2\Delta_a\tilde{\Delta}_a+\Omega^2)$. Third, the
desirable operation $U_{y_1}$ could be constructed as:
$U_{y_1}=\exp(i\pi\sigma_{y_1}/4)=\exp(i\pi\sigma_{z_1}/4)\exp(i3\pi\sigma_{x_1}/4)
\exp(i3\pi\sigma_{z_1}/4)$. It should be pointed out that the
durations $t_x$ (or $t_y$) of the single-qubit operations required
above for implementing the desirable tomographies is estimated as
$\sim100$ps using the experimental parameters:
$\epsilon\sim2\pi\times 20$MHz~\cite{Gambetta}, and $\Delta_{dr}\sim
\kappa/2$~\cite{Blais2}. This is significantly less by at least two
orders than the qubit decoherence time, which is measured as $\sim
500$ns~\cite{Blais2}. Therefore, the required gate operations are
accessible and the proposed tomographic reconstructions are
experimentally feasible.

As an example, we assume that the three parameters $r_x=0.6$,
$r_y=0.5$, $r_z=0.6$ are obtained through the above reconstructions,
then the reconstructed state $\rho_1$ can be written as
$\rho_1=0.8|0\rangle\langle0|
+(0.3-0.25i)|0\rangle\langle1|+(0.3+0.25i)|1\rangle\langle0|+0.2|1\rangle\langle1|$,
whose real $\rho_{ij}^{(R)}$ and imaginary $\rho_{ij}^{(I)}$ parts
(i, j=0, 1) are graphically represented in Fig.~2.

\begin{figure}[htbp]
\includegraphics[width=8cm,height=4cm]{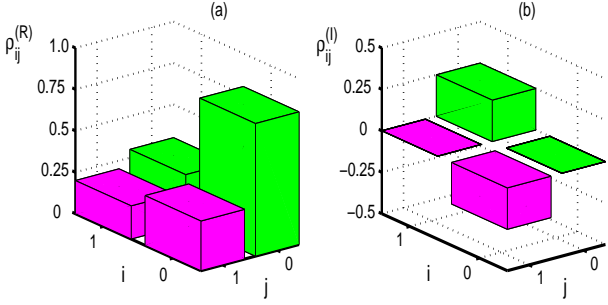}
\caption{(Color online) Graphic representations of the density
matrix $\rho_1$ for a single-qubit state. The real $\rho_{ij}^{(R)}$
and imaginary $\rho_{ij}^{(I)}$ parts of the density matrix elements
$\rho_{ij}=\langle i|\rho|j\rangle$ (i, j=0, 1) are plotted in (a)
and (b), respectively.}
\end{figure}

\section{Tomographic reconstruction of a two-qubit state by
QND measurements}

\subsection{Nondestructive detection of an unknown two-qubit state by cavity transmissions}

We extend the above sing-qubit QND measurements to the two-qubit
case. The transition frequencies of the two qubits are represented
as $\omega_1$ and $\omega_2$, respectively. In the above dispersive
condition (4) and in a framework rotating at $\omega_{d}$, the
effective Hamiltonian of the present complete system is
\begin{eqnarray}
\tilde{H}_2&=&(-\Delta_{dr}+\Gamma_1\sigma_{z_1}+\Gamma_2\sigma_{z_2})\hat{a}^\dagger\hat{a}\nonumber\\
&&+\frac{\tilde{\omega}_1}{2}\sigma_{z_1}+\frac{\tilde{\omega}_2}{2}\sigma_{z_2}
+\epsilon(\hat{a}^\dagger+\hat{a}),
\end{eqnarray}
where $\Gamma_j=g_j^2/\Delta_{j}$ and
$\tilde{\omega}_j=\omega_j+\Gamma_j$, $j=1,2$.

Similarly, the relevant master equation reads
\begin{eqnarray}
\dot{\varrho}_2&=&-i[\tilde{H}_2,\varrho_2]+\kappa \mathcal
{D}[a]\varrho_2+\sum_{j=1,2}\gamma_{1,j} \mathcal {D}[\sigma_{-_j}]\varrho_2\nonumber\\
&&+ \sum_{j=1,2}\frac{\gamma_{\phi,j}}{2}
\mathcal{D}[\sigma_{z_j}]\varrho_2.
\end{eqnarray}
and the equations of motion for the mean values of various
expectable operators are
\begin{subequations}
\label{eq:whole}
\begin{eqnarray}
\frac{d\langle\hat{a}^\dagger\hat{a}\rangle}{dt}&
=&-\kappa{\langle\hat{a}^\dagger\hat{a}\rangle}-2\epsilon\mathrm{Im}{\langle\hat{a}\rangle},\label{subeq:1}
\end{eqnarray}
\begin{eqnarray}
\frac{d\langle\hat{a}\rangle}{dt}&
=&(i\Delta_{dr}-\frac{\kappa}{2}){\langle\hat{a}\rangle}-i\Gamma_1{\langle\hat{a}\sigma_{z_1}\rangle}-i\Gamma_{2}{\langle\hat{a}\sigma_{z_2}\rangle}-i\epsilon,\nonumber\\\label{subeq:2}
\end{eqnarray}
\begin{eqnarray}
\frac{d\langle\hat{a}\sigma_{z_1}\rangle}{dt}&=&(i\Delta_{dr}-\frac{\kappa}{2}-\gamma_{1,1}){\langle\hat{a}\sigma_{z_1}\rangle}-(i\Gamma_1+\gamma_{1,1}){\langle\hat{a}\rangle}\nonumber\\
&&-i\Gamma_2{\langle\hat{a}\sigma_{z_1}\sigma_{z_2}\rangle}-i\epsilon{\langle\sigma_{z_1}\rangle},\label{subeq:3}
\end{eqnarray}
\begin{eqnarray}
\frac{d\langle\hat{a}\sigma_{z_2}\rangle}{dt}&=&(i\Delta_{dr}-\frac{\kappa}{2}-\gamma_{1,2}){\langle\hat{a}\sigma_{z_2}\rangle}-(i\Gamma_2+\gamma_{1,2}){\langle\hat{a}\rangle}\nonumber\\
&&-i\Gamma_{1}{\langle\hat{a}\sigma_{z1}\sigma_{z_2}\rangle}-i\epsilon{\langle\sigma_{z_2}\rangle},\label{subeq:4}
\end{eqnarray}
\begin{eqnarray}
\frac{d\langle\hat{a}\sigma_{z_1}\sigma_{z_2}\rangle}{dt}&=&(i\Delta_{dr}-\frac{\kappa}{2}-\gamma_{1,1}-\gamma_{1,2}){\langle\hat{a}\sigma_{z_1}\sigma_{z_2}\rangle}\nonumber\\
&&-i\epsilon{\langle\sigma_{z_1}\sigma_{z_2}\rangle}-(i\Gamma_2+\gamma_{1,2}){\langle\hat{a}\sigma_{z_1}\rangle}\nonumber\\
&&-(i\Gamma_1+\gamma_{1,1}){\langle\hat{a}\sigma_{z_2}\rangle},\label{subeq:5}
\end{eqnarray}
\begin{eqnarray}
\frac{d\langle\sigma_{z_1}\rangle}{dt}&=&-\gamma_{1,1}({\langle\sigma_{z_1}\rangle}+1),\label{subeq:6}
\end{eqnarray}
\begin{eqnarray}
\frac{d\langle\sigma_{z_2}\rangle}{dt}&=&-\gamma_{1,2}({\langle\sigma_{z_2}\rangle}+1),\label{subeq:7}
\end{eqnarray}
\begin{eqnarray}
\frac{d\langle\sigma_{z_1}\sigma_{z_2}\rangle}{dt}&=&-(\gamma_{1,1}+\gamma_{1,2})\langle\sigma_{z_1}\sigma_{z_2}\rangle-\gamma_{1,1}
\langle\sigma_{z_2}\rangle\nonumber\\
&&-\gamma_{1,2}\langle\sigma_{z_1}\rangle.\label{subeq:8}
\end{eqnarray}
\end{subequations}
Again, due to the relatively-long decoherence times of the qubits
and their sufficiently short measured times, the additional
measurement-induced dephasing and decay rates of the qubits are also
unimportant. Thus, the expectable values of the qubit operators can
still be regarded as unchanged, i.e.,
$\langle\sigma_{z_j}(t)\rangle\approx\langle\sigma_{z_j}(0)\rangle$
and
$\langle\sigma_{z_1}(t)\sigma_{z_2}(t)\rangle\approx\langle\sigma_{z_1}(0)\sigma_{z_2}(0)\rangle$,
during the desirable QND measurements. As a consequence, one can
easily solve the above Eqs. (23a-e) and finally obtain the exact
steady-state distribution of the intracavity photon number
\begin{widetext}
\begin{eqnarray}
\frac{\langle\hat{a}^\dagger\hat{a}\rangle_{ss}}{\epsilon^2}=\frac{2}{\kappa}\mathrm{Re}\left\{\frac{
F(\sum_{j,j'}{B_jD_{j'}G_j}+D_1D_2) +B_1B_2[G_{12}(D_1+D_2)
+\sum_{j,j'}{E_jG_{j'}}]-\sum_{j}{B_jE_j(D_j+B_jG_j)}}
{\sum_{j,j'}B_jE_j(D_{j'}F+D_jA)-{(B_1E_1-B_2E_2)^2}
-AD_1D_2F}\right\},\nonumber\\
 j,j'=1,2,
j\neq{j'}.
\end{eqnarray}
\end{widetext}
Here,
$A=i\Delta_{dr}-\frac{\kappa}{2},B_j=i\Gamma_j,D_j=i\Delta_{dr}-\frac{\kappa}{2}-\gamma_{1,j},
E_j=i\Gamma_j+\gamma_{1,j},F=i\Delta_{dr}-\frac{\kappa}{2}-\gamma_{1,j}-\gamma_{1,j'}$,
$G_j=\langle\sigma_{z_j}(0)\rangle$, and
$G_{12}=\langle\sigma_{z_1}(0)\sigma_{z_2}(0)\rangle$.

  \begin{figure}[htbp]
\includegraphics[width=8cm,height=6cm]{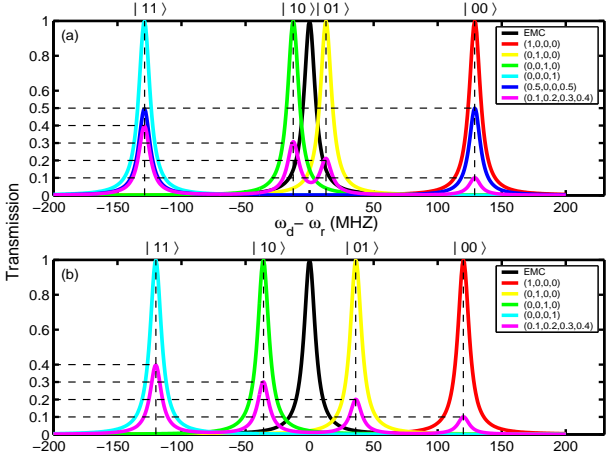}
\caption{(Color online) (a). Cavity transmission of the cavity
versus the probe detuning $\omega_d-\omega_r$ for certain slected
two-qubit states with $|\alpha_0|^2=1$, $|\alpha_1|^2=1$,
$|\alpha_2|^2=1$, $|\alpha_3|^2=1$, $(|\alpha_0|^2$, $|\alpha_1|^2$,
$|\alpha_2|^2$, $|\alpha_3|^2)=$ $(0.5, 0, 0, 0.5)$,  and
$(|\alpha_0|^2$, $|\alpha_1|^2$, $|\alpha_2|^2$, $|\alpha_3|^2)=$
$(0.1, 0.2, 0.3, 0.4)$, respectively. For comparison, the empty
cavity (EMC) transmission is also plotted in black line. Here,
 The parameters are
$(\Gamma_1,\Gamma_2,\kappa,\gamma_{1,1},\gamma_{1,2})
 =2\pi\times(-11.11,-9.11,1.7,0.02,0.022,)$MHz
~\cite{Filipp,Wallraff2}. In (b) only the parameters $\Gamma_1$ and
$\Gamma_2$ are modified as $\Gamma_1'=1.05 \Gamma_1$ and
$\Gamma_2'=0.85\Gamma_2$. In this case, the relative heights of the
peaks are exactly equivalent to the corresponding probabilities of
the single logic states superposed in the measured superposition
state.}
\end{figure}

We now investigate the above distributions schematically for various
typically selected two-qubit initial states. First, we assume that
the two-qubit is initially prepared at only one of the four logic
states, i.e., in the generic expression
$|\psi\rangle_2=\alpha_1|00\rangle+\alpha_2|01\rangle+\alpha_3|10\rangle+\alpha_4|11\rangle$
only one of the four probability amplitudes equals $1$, e.g.,
$\alpha_1=1$, $\alpha_2=\alpha_3$ $=\alpha_4$ $=0$. Fig.~3 shows
clearly that single peaks reveal the inputs of these four single
logic states, and they can also be distinguished by the shifts of
the central frequencies of the transmission spectrum. The peaks with
frequency shifts: $-\Gamma_1-\Gamma_2$, $-\Gamma_1+\Gamma_2$,
$\Gamma_1-\Gamma_2$, and $\Gamma_1+\Gamma_2$ mark the state
$|00\rangle$, $|01\rangle$, $|10\rangle$, and $|11\rangle$,
respectively. Thus the pulls of the cavity are strongly dependent of
the states of the qubits. For these single logic states the heights
of the single peaks are exactly equivalent and of unity value, which
is the same as that for the EMC. Next, for the superposition of the
four single logic states the situations are quite different. For
example, Fig.~3 (a) also shows that, if the two-qubit is prepared
initially as one of the Bell states: $(|\alpha_1|^2$,
$|\alpha_2|^2$, $|\alpha_3|^2$, $|\alpha_4|^2)=(0.5, 0, 0, 0.5)$,
then the transmitted spectrum of the cavity reveals two peaks; their
locations are respectively at the positions for the single states
$|00\rangle$ and $|11\rangle$, but have the same relative heights.
Moreover, for a more generic superposed state $(|\alpha_1|^2$,
$|\alpha_2|^2$, $|\alpha_3|^2$, $|\alpha_4|^2)=(0.1, 0.2, 0.3, 0.4)$
one can see that four peaks are exhibited simultaneously. The
central positions of these peaks locate at the corresponding
positions of single logic states $|00\rangle, |01\rangle,
|10\rangle$, and $|11\rangle$, respectively. The relative heights of
them read $0.1, 0.212, 0.308$, and $0.4$, respectively. Here, the
relative heights of the peaks marking the states $|00\rangle$ and
$|11\rangle$ are exactly equivalent to the superposed probabilities
$|\alpha_1|^2$ and $|\alpha_4|^2$. However, the relative heights of
the peaks marking the states $|01\rangle$ and $|10\rangle$ deviate
from the corresponding superposed probabilities $|\alpha_2|^2$ and
$|\alpha_3|^2$. This is because these two peaks are not well
distinguished due to the contributions from these two logic states'
overlap. As a consequence, each peak is higher a little than the
expected one, i.e., the superposed probability of the relevant logic
state. While, such a situation does not exist for the $|00\rangle$
and $|11\rangle$ peaks (the relative heights of them equal to the
expected ones), as they are separated sufficiently far from the
others. In Fig.~3 (b) we modify the relevant parameters such as
$\Gamma_1'=1.05 \Gamma_1$ and $\Gamma_2'=0.85\Gamma_2$. Then we find
that each peak of the transmission of the cavity is well separated
from the others, and thus its relative height is exactly equal to
the expectable superposed probability of the corresponding logic
state in the measured two-qubit state.

\subsection{High efficiency tomographic reconstruction of a two-qubit state}

The two-qubit state tomography is done in the same way as that for
the single-qubit state. The only difference is  that now there are
$15$ real parameters to be determined in the 4-dimensional density
matrix operator $\rho_2$, and thus more operations are required to
transfer the nondiagonal elements in $\rho_2$ to the diagonal
locations. Generally, the 4-dimensional density matrix for a
two-qubit state
$|\psi\rangle_2=\alpha_1|00\rangle+\alpha_2|01\rangle+\alpha_3|10\rangle+\alpha_4|11\rangle$
in a complete basis $\{|1\rangle_2=|00\rangle$,
$|2\rangle_2=|01\rangle$, $|3\rangle_2=|10\rangle$,
$|4\rangle_2=|11\rangle\}$ can be represented as
\begin{eqnarray}
 \rho_2=\left(
\begin{array}{cccc}
\rho_{11} & \rho_{12}&\rho_{13} & \rho_{14}  \\
\rho_{21} & \rho
_{22}&\rho_{23} & \rho_{24}  \\
\rho_{31} & \rho
_{32}&\rho_{33} & \rho_{34}  \\
\rho_{41} & \rho_{42}&\rho_{43} & \rho_{44}  \\
\end{array}\right),
\end{eqnarray}
which can also be rewritten as $\rho_2=\bar{\rho}_2/4$
with~\cite{Liu}
\begin{widetext}
\begin{eqnarray}
&& \bar{\rho}_2=\sum_{m,n=0,x,y,z}r_{mn}\sigma_{m_1}\otimes\sigma_{n_2}\nonumber\\
&&=\left(
\begin{array}{cccc}
 r_{00}+r_{0z}+r_{z0}+r_{zz} &r_{0x}+r_{zx}-ir_{0y}-ir_{zy} &r_{x0}+r_{xz}-ir_{y0}-ir_{yz} &r_{xx}-r_{yy}-ir_{xy}-ir_{yx} \\
 r_{0x}+r_{zx}+ir_{0y}+ir_{zy} &r_{00}-r_{0z}+r_{z0}-r_{zz} &r_{xx}+r_{yy}+ir_{xy}-ir_{yx} &r_{x0}-r_{xz}-ir_{y0}+ir_{yz} \\
r_{x0}+r_{xz}+ir_{y0}+ir_{yz} &r_{xx}+r_{yy}-ir_{xy}+ir_{yx} &r_{00}+r_{0z}-r_{z0}-r_{zz} &r_{0x}-r_{zx}-ir_{0y}+ir_{zy} \\
r_{xx}-r_{yy}+ir_{xy}+ir_{yx} &r_{x0}-r_{xz}+ir_{y0}-ir_{yz} &r_{0x}-r_{zx}+ir_{0y}+ir_{zy} &r_{00}-r_{0z}-r_{z0}+r_{zz}\\
\end{array}\right).\nonumber\\
\end{eqnarray}
\end{widetext}
Here, $\sigma_{m=x,y,z}$ are the Pauli operators and $\sigma_{0}$
identity matrix, and what we want to determine is sixteen real
parameters $r_{mn}$. Note that the first and second subscripts of
the matrix elements $\rho_{ij}$ (i, j=1, 2, 3, 4) in  Eq.~(25) and
$r_{mn}$ in Eq.~(26) is labeled for the first and second qubits,
respectively.

As in the above discussion, performing the QND measurements on the
two-qubit state $\rho_2$ can directly determine all the four
diagonal elements: $\rho_{11}$, $\rho_{22}$, $\rho_{33}$ and
$\rho_{44}$, respectively, by the measured results $|\alpha_1|^2$,
$|\alpha_2|^2$, $|\alpha_3|^2$ and $|\alpha_4|^2$. As a consequence,
the parameters $r_{00}$, $r_{0z}$, $r_{z0}$ and $r_{zz}$ can be
determined by
\begin{eqnarray}
 &&r_{00}=|\alpha_1|^2+|\alpha_2|^2+|\alpha_3|^2+|\alpha_4|^2=1,\nonumber\\
&&r_{0z}=|\alpha_1|^2-|\alpha_2|^2+|\alpha_3|^2-|\alpha_4|^2 ,\nonumber\\
&&r_{z0}=|\alpha_1|^2+|\alpha_2|^2-|\alpha_3|^2-|\alpha_4|^2,\nonumber\\
&&r_{zz}=|\alpha_1|^2-|\alpha_2|^2-|\alpha_3|^2+|\alpha_4|^2.
\end{eqnarray}
To determine the other $12$ parameters, we need to perform certain
unitary operations to transfer them to the diagonal locations for
other QND measurements.

It is well-known that, arbitrary two-qubit operation assisted by
arbitrary rotations of the single qubits generate an universal set
of quantum gates. So the key to implement the above required
operations for tomographies is to realize a two-qubit gate. Again,
for the experimental circuit QED system with two superconducting
charge qubits, such a gate could be implemented by using the
so-called FLICFORQ protocol~\cite{Blais2}. In fact, if the cavity is
driven by two external fields satisfying the sideband matching
condition: $\omega_{d_2}-\omega_{d_1}=\Omega_1+\Omega_2$, then an
effective Hamiltonian
\begin{eqnarray}
\tilde{H}_{\rm{FF}}=\omega_r\hat{a}^\dagger\hat{a}+
\frac{g_1g_2(\Delta'_1+\Delta'_2)}{16\Delta'_1\Delta'_2}(\sigma_{y_1}\otimes\sigma_{y_2}+
\sigma_{z_1}\otimes\sigma_{z_2}),\nonumber\\
\end{eqnarray}
can be induced in a quadruply rotating framework. Here,
$\Delta'_j={\omega}_{_j}+2{\Omega^2_{jj'}}/{\Delta_{_{jdj'}}}-\omega_r$
with $\Omega_{jj'}=2g_j\epsilon_{j'}/(\omega_{d_{j'}}-\omega_r)$,
and $\Delta_{_{jdj'}}=\omega_{_j}-\omega_{d_{j'}}$, $j, j'=1, 2$, $
j\neq j'$. Obviously, the evolution under the above Hamiltonian with
the duration, e.g., around $\sim100$ps, for the experimental
parameters~\cite{Blais2}, can produce a two-qubit operation:
\begin{equation}
U_{\rm{FF}}=\exp[i{\pi}(\sigma_{y_1}\otimes\sigma_{y_2}+
\sigma_{z_1}\otimes\sigma_{z_2})/{4}].
\end{equation}
On the other hand, the typical single-qubit gates, e.g.,
$U_{x_j},\,U_{y_j}$, and $U_{z_j}$ (j=1, 2) can be relatively easy
to produce using the similar approaches presented in Sec. III. With
such a two-qubit operation and these single-qubit gates, we show in
Table I how to perform the desirable unitary operations for
transferring the non-diagonal elements to the diagonal locations.
For example, by performing a selected operational sequence
$W=U_{\rm{FF}}U_{x_1}$ on the original density matrix $\rho_2$, we
have a new density matrix $\rho_2'=W \rho_2 W^\dagger$, and the new
diagonal elements are
\begin{eqnarray}
  &&\rho_{11}'=\frac{1}{4}(r_{00}+r_{xy}-r_{yz}+r_{zx}),\nonumber\\
 &&\rho_{22}'=\frac{1}{4}(r_{00}+r_{xy}+r_{yz}-r_{zx}),\nonumber\\
&&\rho_{33}'= \frac{1}{4}( r_{00}-r_{xy}+r_{yz}+r_{zx}),\nonumber\\
&&\rho_{44}'=\frac{1}{4}( r_{00}-r_{xy}-r_{yz}-r_{zx}).
\end{eqnarray}
Then, by the QND measurements the values of
$|\alpha'_1|^2,|\alpha'_2|^2$, $|\alpha'_3|^2$, and $|\alpha'_4|^2$
are given directly. As a consequence, the desirable parameters
$r_{00}$, $r_{xy}$, $r_{yz}$ and $r_{zx}$ can be obtained by the
relations:
\begin{eqnarray}
 &&r_{00}=|\alpha'_1|^2+|\alpha'_2|^2+|\alpha'_3|^2+
|\alpha'_4|^2=1,\nonumber\\
&&r_{xy}=|\alpha'_1|^2+|\alpha'_2|^2-|\alpha'_3|^2-|\alpha'_4|^2,\nonumber\\
&&r_{yz}=-|\alpha'_1|^2+|\alpha'_2|^2+|\alpha'_3|^2-|\alpha'_4|^2,\nonumber\\
&&r_{zx}=|\alpha'_1|^2-|\alpha'_2|^2+|\alpha'_3|^2-|\alpha'_4|^2.
\end{eqnarray}
Similarly, other non-diagonal elements can also be determined.
Clearly, here only six kinds of QND measurements are sufficient to
tomographically reconstruct a two-qubit state. This is obviously
simpler than the previous tomographies based on the usual DP
measurements, wherein $15$ kinds of measurements are probably
required~\cite{Liu,Filipp}. Thus, the present tomographies is
essentially high efficient.

After performing all the QND measurements listed in the table, a
two-qubit state can be completely reconstructed. For example, a
two-qubit state $\rho_2$ having the following representation:
\begin{eqnarray}
  \rho_2=\left(
\begin{array}{cccc}
 r_{00}&r_{0x}&r_{0y}&r_{0z}\\
 r_{x0}&r_{xx}&r_{xy}&r_{xz}\\
r_{y0}&r_{yx}&r_{yy}&r_{yz}\\
r_{z0}&r_{zx}&r_{zy}&r_{zz}\\
\end{array}\right)=
\left(
\begin{array}{cccc}
  1 & 0 & 0 & -\frac{1}{5} \\
 0 & \frac{1}{4} & 0 & \frac{3}{5} \\
 0 & 0 & -\frac{1}{4} & 0 \\
 -\frac{2}{5} & \frac{1}{8} & 0 & 0 \\
\end{array}\right),
\end{eqnarray}
which can be effectively reconstructed by these parameters
\begin{widetext}
\begin{eqnarray}
 \rho_2=\left(
\begin{array}{cccc}
\rho_{11} & \rho_{12}&\rho_{13} & \rho_{14}  \\
\rho_{21} & \rho
_{22}&\rho_{23} & \rho_{24}  \\
\rho_{31} & \rho
_{32}&\rho_{33} & \rho_{34}  \\
\rho_{41} & \rho_{42}&\rho_{43} & \rho_{44}  \\
\end{array}\right)
=\left(
\begin{array}{cccc}
0.1 & 0.0313-0.0313i& 0.15-0.15i & -0.125i \\
  0.0313+0.0313i&   0.2 & 0.125 & -0.15+0.15i\\
0.15+ 0.15i & 0.125 & 0.3 & -0.0313+0.0313i\\
 0.125i& -0.15-0.15i & -0.0313+0.0313i & 0.4\\
\end{array}\right),
\end{eqnarray}
\end{widetext}
determined by six kinds of QND measurements. The simulated
reconstructions are graphically shown in Fig.~4, where
$\rho_{ij}^{(R)}$ and $\rho_{ij}^{(I)}$ are the real and imaginary
parts of the reconstructed state $\rho_2$ in the bases
$|1\rangle_2=|00\rangle$, $|2\rangle_2=|01\rangle$,
$|3\rangle_2=|10\rangle$, $|4\rangle_2=|11\rangle$, with $i, j=1, 2,
3, 4$.

  \begin{figure}[htbp]
\includegraphics[width=8cm,height=4cm]{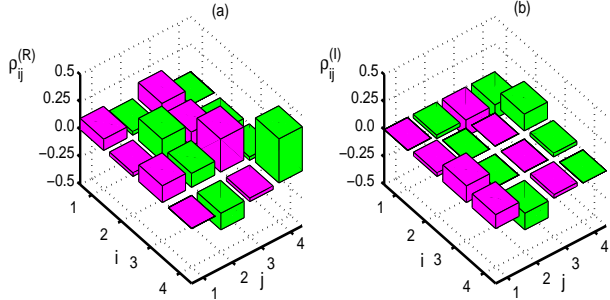}
\caption{(Color online) Schematic representations of the density
matrix $\rho_2$ for a two-qubit state. The real $\rho_{ij}^{(R)}$
and imaginary $\rho_{ij}^{(I)}$ parts $(i, j=1, 2, 3, 4)$ of the
density matrix elements
 in the complete bases are plotted in (a) and
(b), respectively.}
\end{figure}

\begin{table}
\caption{The operational combinations before the QND measurements to
determine the parameters for tomographically reconstructing a
two-qubit state. The subscript $"1(2)"$ of $U$ is labeled for the
operation of qubit $1(2)$.}
\begin{ruledtabular}
\begin{tabular}{lcr}
quantum operation $W$ &  determined parameters\\
\hline
no &  $r_{00}$, $r_{zz}$, $r_{0z}$, $r_{z0}$\\
$U_{\rm{FF}}U_{x_1}$ &  $r_{00}$, $r_{xy}$, $r_{yz}$, $r_{zx}$\\
$U_{\rm{FF}}U_{y_1}$ &  $r_{00}$, $r_{yx}$, $r_{zy}$, $r_{xz}$\\
$U_{\rm{FF}}U_{z_1}$ &  $r_{00}$, $r_{xx}$, $r_{yy}$, $r_{zz}$\\
$U_{y_1}U_{z_1}U_{\rm{FF}}$ &  $r_{00}$, $r_{0x}$, $r_{y0}$, $r_{yx}$\\
$U_{y_2}U_{z_2}U_{\rm{FF}}$ &  $r_{00}$, $r_{x0}$, $r_{0y}$, $r_{xy}$\\
\end{tabular}
\end{ruledtabular}
\end{table}

\section{Discussions and Conclusions}

Generally, the quantum state tomographic constructions demonstrated
above can be extended for $N$ (with $N>2$) qubits in a
straightforward manner. This is because that the proposed QND
measurements can be directly applied to determine all the diagonal
elements of the arbitrary $N$-qubit state; the individual superposed
logic states can be inferred from the relevant positions of the
measured peaks, and the probabilities of the corresponding
computational bases superposed in the measured state could be
extracted from the relative heights of the peaks (when they separate
sufficiently from the others). Moreover, all the required operations
for the tomographic reconstructions can be implemented from the
universal set of the logic gates demonstrated.

In summary, we have proposed a scheme to perform the quantum state
tomographies by QND measurements. Differing from the usual
tomographies based on the DP measurements, here the QND measurements
are utilized. Since all the diagonal elements of the density matrix
of an unknown quantum state can be simultaneously determined by a
single kind of QND measurements, the efficiency of the present
tomographic reconstruction is definitely better for more qubits.
Specifically, our proposal is demonstrated with the current circuit
QED setup with a few charge qubits, and could be generalized to
other systems, in principle.


\section*{Acknowledgments}

This work was supported in part by the National Science Foundation
grant No. 10874142, 90921010, and the National Fundamental Research
Program of China through Grant No. 2010CB923104, and the Fundamental
Research Funds for the Central Universities No. SWJTU09CX078, and
A*STAR of Singapore under research grant No. WBS: R-144-000-189-305.

\end{document}